\documentclass[aps,pra,twocolumn,showpacs,superscriptaddress]{revtex4-1}

\usepackage{amsfonts,amssymb,amsmath}
\usepackage{mathtools}
\usepackage[]{graphics,graphicx,epsfig}
\usepackage{amsthm,multirow}
\usepackage{color}
\begin{document}

\title{Duality in entanglement of macroscopic states of light}

\author{Su-Yong \surname{Lee}}
\affiliation{School of Computational Sciences, Korea Institute for Advanced Study, Hoegi-ro 85,Dongdaemun-gu, Seoul 02455, Korea}

\author{Chang-Woo Lee}
\affiliation{School of Computational Sciences, Korea Institute for Advanced Study, Hoegi-ro 85,Dongdaemun-gu, Seoul 02455, Korea}

\author{Pawe\l{} Kurzy\'nski}    \affiliation{Faculty of Physics, Adam Mickiewicz University, Umultowska 85, 61-614 Pozna\'n, Poland}  \affiliation{Centre for Quantum Technologies, National University of Singapore, 3 Science Drive 2, Singapore 117543}

\author{Dagomir Kaszlikowski}   \affiliation{Centre for Quantum Technologies, National University of Singapore, 3 Science Drive 2, Singapore 117543} \affiliation{Department of Physics, National University of Singapore, 2 Science Drive 3, Singapore 117542 }

\author{Jaewan Kim}
\affiliation{School of Computational Sciences, Korea Institute for Advanced Study, Hoegi-ro 85,Dongdaemun-gu, Seoul 02455, Korea}

\date{\today}

\begin{abstract}
We investigate duality in entanglement of a bipartite multi-photon system generated from a coherent state of light. The system can exhibit polarization entanglement if the two parts are distinguished by their parity, or parity entanglement if the parts are distinguished by polarization. It was shown in [PRL 110, 140404 (2013)] that this phenomenon can be exploited as a method to test indistinguishability of two particles and it was conjectured that one can also test indistinguishability of macroscopic systems. We propose a setup to test this conjecture. Contrary to the previous studies using two-particle interference effect as in Hong-Ou- Mandel setup, our setup neither assumes that the tested state is composed of single particles nor requires that the total number of particles be fixed. Consequently the notion of entanglement duality is shown to be compatible with a broader class of physical systems. Moreover, by observing duality in entanglement in the above system one can confirm that macroscopic systems exhibit quantum behaviour. As a practical side, entanglement duality is a useful concept that enables adaptive conversion of entanglement of one degree of freedom (DOF) to that of another DOF according to varying quantum protocols.
\end{abstract}

\maketitle

\section{Introduction}


 In the scenario of quantum systems, quantum indistinguishability of identical particles has been tested using the interference effect of a Hong-Ou-Mandel-type scheme \cite{HOM87}. It assumes that indistinguishable particles scatter independently, e.g., on a beam splitter and do not interact. One cannot, however, exclude the possibility that the resultant bunching or anti-bunching effect actually originates from the interaction between distinguishable particles.
Therefore, in order to verify true particle indistinguishability one needs to 
preclude the possibility of inter-particle interaction. 
One such way is to prepare an entangled state of identical particles and probe if the entanglement encoded in a certain degree of freedom (DOF) can be converted to one in another DOF.
This interchangeability of entanglement between different DOFs is studied in Ref. \cite{BH13}
and dubbed ``duality in entanglement.''
Since such duality does not arise in case of distinguishable (e.g., different species of) particles,
it is considered a novel way to manifest quantum indistinguishability.
 For example, if two single photons are generated in a parametric down-conversion and thereby are entangled in polarization DOF ($H,V$), the entanglement can be accessed because the two particles are effectively distinguishable by their path DOF, say, (1, 2). However, if one decides to effectively distinguish the identical particles by their polarizations, one will observe entanglement in the path DOF. This phenomenon would not occur for distinguishable particles and hence it can be used in testing their indistinguishability. 

To date, such tests of quantum indistinguishability based on entanglement of two identical particles were performed in two scenarios: the first case utilizes a polarization/path entangled state \cite{BH13} and the duality is tested by the violation of a Bell's inequality; the second case considers a spin/orbital angular momentum entangled state \cite{BZA15} and the duality is tested by an entanglement witness. 
Both the scenarios are implemented on photonic setup and the corresponding demonstrations remain only at single-photon level.
Therefore, one research direction in indistinguishability test based on entanglement duality is to survey schemes that choose other kinds of identical particles (e.g., other bosonic particles or fermions); 
another direction is to examine if the notion of entanglement duality can also apply to a multi-particle system beyond the aforementioned single-particle level.
We are interested in the latter direction and here propose a scenario for entanglement duality test in a macroscopic bipartite system which surpasses single-particle level and even does not have a fixed number of particles. Note that this situation differs from the Hong-Ou-Mandel scenario in which one tests distinguishability of only two microscopic particles.

An additional motivation for our studies comes from the fact that it was conjectured in \cite{BH13} that entanglement duality scenario can be used to test indistinguishability of complex macroscopic objects. Such macroscopic system have many internal degrees of freedom and two macro-molecules that are initially identical will most probably evolve in a different way leading to effective distinguishability. In this case duality in entanglement would be unobservable. This can be considered as a kind of transition from quantum to classical domain. However, if duality in entanglement occurs for macroscopic objects, then we can confirm that despite large size the systems are still quantum since they exhibit indistinguishability that is a truly quantum feature.

We consider macroscopic light field states that are entangled in polarization DOF and by entanglement duality can also be regarded as entangled in parity DOF. 
Specifically, we consider coherent states---in principle their size can be arbitrarily large---which can be effectively distinguished by parity (the former case) or by polarization (the latter):
\begin{eqnarray}
&&\frac{1}{\sqrt{2}}(|H\rangle_{even}|V\rangle_{odd}\pm|V\rangle_{even}|H\rangle_{odd})\nonumber\\
&& = \frac{1}{\sqrt{2}}(|even\rangle_H|odd\rangle_V\pm|odd\rangle_H|even\rangle_V).
\label{eq:entangledstate1}
\end{eqnarray}
Here,  $|even\rangle=N_e(|\alpha\rangle+|-\alpha\rangle)$ and $|odd\rangle=N_o(|\alpha\rangle-|-\alpha\rangle)$ are even and odd coherent states with normalization constants $N_e$ and $N_o$ respectively. 
We adopt orthogonal coherent state basis $\{|even\rangle,~|odd\rangle \}$ instead of non-orthogonal one $\{|\alpha\rangle,~|-\alpha\rangle \}$. 
Notice that an even coherent state is orthogonal to an odd coherent state since an even (odd) coherent state is a superposition of even- (odd-) number Fock states. The entanglement of (macroscopic) coherent states encoded in polarization/parity DOF is interchangeable between these two DOFs and accordingly its duality in entanglement can be identified. 

To obtain the entangled states in Eq. \eqref{eq:entangledstate1}, it is required first to prepare a single-mode even (odd) coherent state which is experimentally within reach in a trapped $^9$Be$^+$ ion system \cite{MMKW96}, a high \textit{Q} microwave cavity \cite{B96}, a Bose-Einstein condensate with Rb atoms \cite{GMHB02}, and an optical system using homodyne detection and Fock states \cite{OJTG07}.
To identify entanglement in each DOF, one can consider quantum information protocols such as CHSH-Bell-type inequality test based on displaced parity detector \cite{BW99} or interaction-free measurement scheme \cite{EV93}.
Furthermore, instead of using the definite-parity coherent states one can consider another macroscopic state basis, namely squeezed vacuum and a single-photon-subtracted squeezed vacuum state, which also comprises a parity-based orthogonal basis:
Squeezed vacuum state is a superposition of even-number Fock states whereas its single-photon-subtracted version consists of odd-number Fock states.

This paper is organized as follows. We begin with generation scheme of the entangled states in Eq. \eqref{eq:entangledstate1}. Then, we discuss how to identify entanglement in each DOF. Next, we discuss a similar scenario using squeezed vacuum states. Finally, we summarize our results and list open questions.

\section{State generation scheme}


\begin{figure}
\centerline{\scalebox{0.35}{\includegraphics[angle=0]{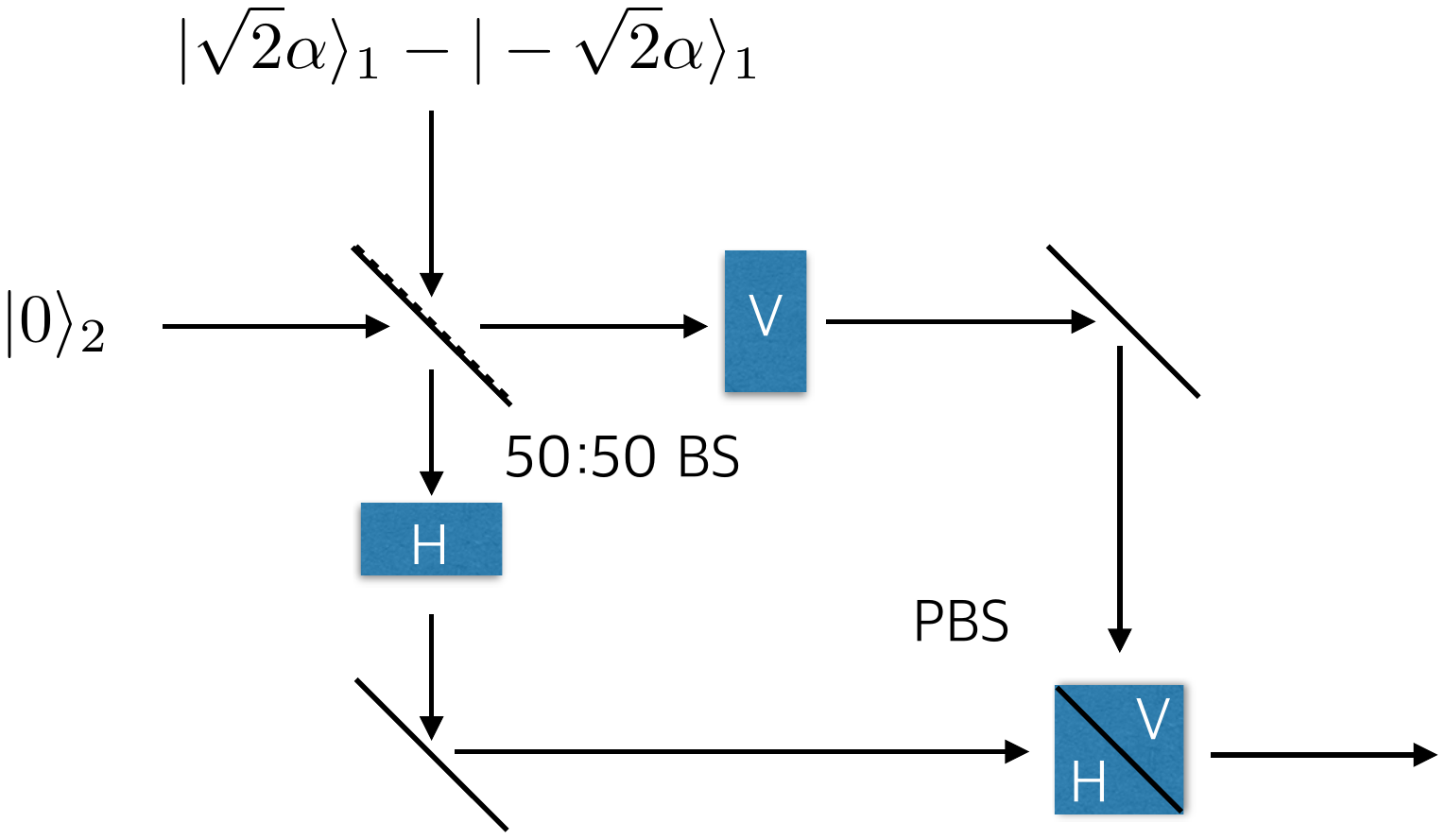}}}
\vspace{-1.6in}
\caption{State generation scheme of Eq. (1). H (V) represents H- (V-) polarizer.
PBS is a polarizing beam splitter which transmits horizontal polarization and reflects vertical polarization.
}
\label{fig:fig1}
\end{figure}

The desired entangled states in Eq. \eqref{eq:entangledstate1} can be generated by injecting an odd coherent state into a 50:50 beam splitter as in Fig. \ref{fig:fig1}.
The output state is simply reformulated in an orthonormal basis $\{|even\rangle,~|odd\rangle \}$ and it is given by
\begin{eqnarray}
	&&\hat{B}_{ab}(|\sqrt{2}\alpha\rangle_1-|-\sqrt{2}\alpha\rangle_1)|0\rangle_2\nonumber\\
	&&=|\alpha\rangle_1|-\alpha\rangle_2-|-\alpha\rangle_1|\alpha\rangle_2\nonumber\\
	&&\approx |even\rangle_1|odd\rangle_2-|odd\rangle_1|even\rangle_2.
\end{eqnarray}
Notice that if we adjust the phase of a beam splitting operator we can also obtain
$|even\rangle_1|odd\rangle_2 + |odd\rangle_1|even\rangle_2$ from an odd coherent state. 
After applying H- (V-) polarizer into path-mode $1$ ($2$), the entangled state can be represented by
\begin{eqnarray}
	|even\rangle_{H,1}|odd\rangle_{V,2}-|odd\rangle_{H,1}|even\rangle_{V,2}. 
\end{eqnarray}
Now combining both path modes with a polarizing beam splitter, we obtain one of the entangled states in Eq. \eqref{eq:entangledstate1},
\begin{eqnarray}
	|\psi\rangle&\equiv&\frac{1}{\sqrt{2}}(|even\rangle_H|odd\rangle_V-|odd\rangle_H|even\rangle_V)\nonumber\\
	&=&\frac{1}{\sqrt{2}}(|H\rangle_{even}|V\rangle_{odd}-|V\rangle_{even}|H\rangle_{odd}), 
	\label{eq:entangledstate2}
\end{eqnarray}
where the entangled state is on a single path-mode ($1$ or $2$).
Note that we cannot produce the dual entanglement with an even coherent state since it is not interchangeable between the two DOFs.
If one uses an even coherent state in Fig. \ref{fig:fig1}, an entangled coherent state $|\alpha\rangle_1|- \alpha\rangle_2 + |- \alpha\rangle_1| \alpha\rangle_2$ or $|\alpha\rangle_1| \alpha\rangle_2 + |- \alpha\rangle_1| -\alpha\rangle_2$ will be obtained, which does not have a full \textit{ebit} and hence cannot be converted to a maximally entangled polarization state as \eqref{eq:entangledstate2}.
Put in more detail, the final state would be 
$(1/N_e^2) |even\rangle_H|even\rangle_V \pm (1/N_o^2) |odd\rangle_H|odd\rangle_V $ which cannot be converted to a coefficient-balanced entangled state as \eqref{eq:entangledstate2}.

\section{Accessing Entanglement}

\begin{figure}
\centerline{\scalebox{0.31}{\includegraphics[angle=0]{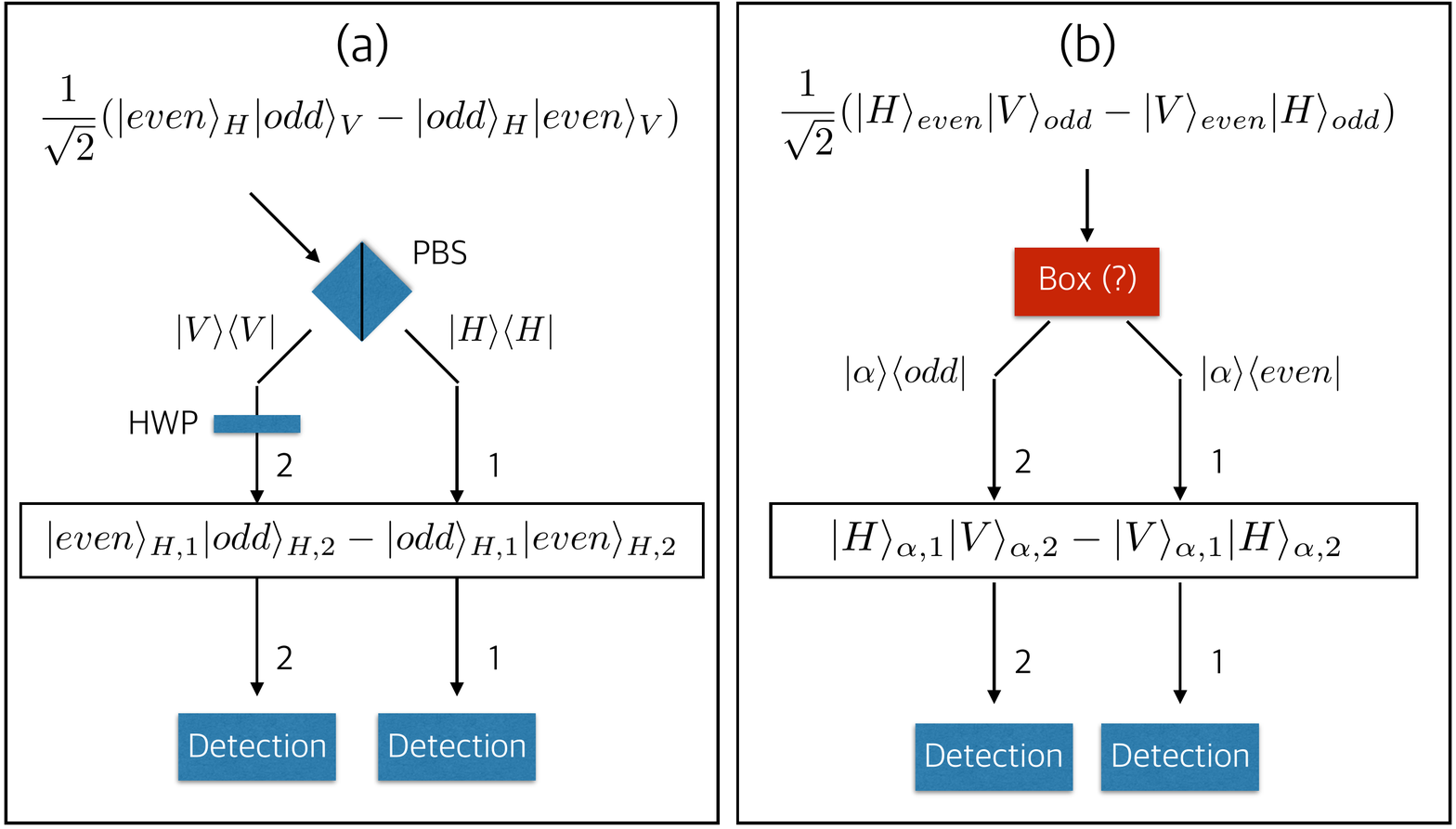}}}
\vspace{-0.6in}
\caption{Detecting entanglement in each degree of freedom (DOF). (a) Parity entanglement. (b) Polarization entanglement. PBS is a polarizing beam splitter which transmits horizontal polarization and reflects vertical polarization. HWP is a half-wave plate which changes polarization from $V$ to $H$ and vice versa.
}
\label{fig:fig2}
\end{figure}

Now we illustrate how to access entanglement in Eq. \eqref{eq:entangledstate2} by directing the two mode variables of DOF to different path modes; see Fig. \ref{fig:fig2}.
Note that placing detectors at both ends as in the figure allows for testing the duality in entanglement;
different types of detections in each DOF setup verify the duality. 
Alternatively, the duality can be tested indirectly by using these path-divided entangled states in disparate information protocols, which will be presented in the next section.

To observe entanglement in parity DOF one just needs a polarizing beam splitter (PBS) and a half-wave plate ahead of detections, as shown in Fig. \ref{fig:fig2}(a).
For the case of polarization-DOF entanglement, however, the procedure is more complicated, and we denote it by the box in Fig. \ref{fig:fig2}(b). 
In the rest of this section, we elaborate on what kinds of processes are included in the box. 

\begin{figure}
\centerline{\scalebox{0.3}{\includegraphics[angle=0]{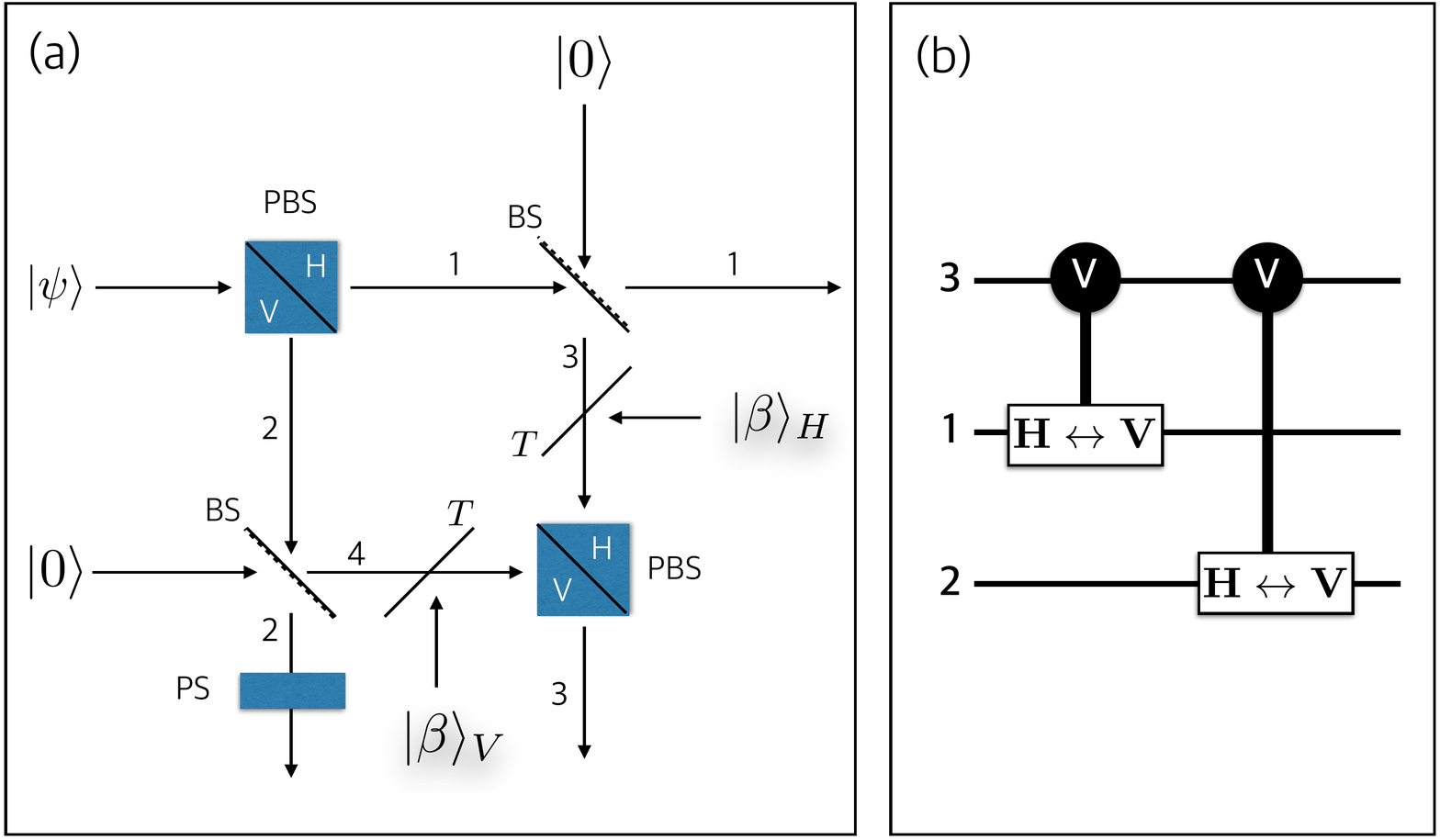}}}
\vspace{-0.6in}
\caption{Box of Fig. 2 (b) to observe polarization entanglement. (a) First, an additional mode $3$ is added to control target modes $1$ and $2$, where $|\psi\rangle=\frac{1}{\sqrt{2}}(|H\rangle_{even}|V\rangle_{odd}-|V\rangle_{even}|H\rangle_{odd})$. 
Displacement operation ($|\beta\sqrt{1-T}\rangle$) is achieved with strong coherent light ($|\beta\rangle$) and a beam splitter with high transmittance ($T\sim 1$) \cite{P96,LB02} .
(b) Next, two controlled NOT gates flip the polarization of the target modes $1$ and $2$ if the control mode $3$ is vertically polarized. 
}
\label{fig:fig3}
\end{figure}

As can be hinted from the two parts of Fig. \ref{fig:fig3}, the box is composed of two stages.
First, we introduce an additional mode $3$ to control the target modes 1 and 2, as shown in Fig. \ref{fig:fig3}(a). The entangled state of Eq. \eqref{eq:entangledstate2} passes through a PBS such that the horizontal (vertical) polarization state moves to path mode $1$ ($2$). Then, impinging each path mode on a 50:50 beam splitter with additional modes $3$ and $4$ which are in vacuum state, we get
\begin{eqnarray}
&&\frac{1}{\sqrt{2}}(|A\rangle_{H,1}|\!-\!A\rangle_{V,2}|A\rangle_{H,3}|\!-\!A\rangle_{V,4}\nonumber\\
&&-|\!-\!A\rangle_{H,1}|A\rangle_{V,2}|\!-\!A\rangle_{H,3}|A\rangle_{V,4}),
\end{eqnarray}
where $A=\alpha/\sqrt{2}$.
After applying displacement operations $\hat{D}_{3,H}(A)$ and $\hat{D}_{4,V}(A)$ to modes $3$ and $4$ respectively, and guiding those modes into another PBS, we obtain the following state:
\begin{eqnarray}
\frac{1}{\sqrt{2}}(|A\rangle_{H,1}|A\rangle_{V,2}|2A\rangle_{H,3}-
|\!-\!\!A\rangle_{H,1}|\!-\!\!A\rangle_{V,2}|2A\rangle_{V,3}),\nonumber\\
\end{eqnarray}
where a phase-shift operation $e^{i\pi\hat{a}^{\dag}_{2,V}\hat{a}^{\phantom{\dag}}_{2,V}}$ was applied to the mode $2$.

Now we are ready to manipulate the target modes $1$ and $2$ under the control mode $3$.
That is, their polarizations are flipped if the control mode $3$ is vertically polarized, as shown symbolically in Fig. \ref{fig:fig3}(b),
\begin{eqnarray}
\frac{1}{\sqrt{2}}(|A\rangle_{H,1}|A\rangle_{V,2}|2A\rangle_{H,3}-
|\!-\!A\rangle_{V,1}|\!-\!A\rangle_{H,2}|2A\rangle_{V,3}).\nonumber\\
\end{eqnarray}
It is implemented by two controlled-NOT-type gates that were realized in optical frequency regime \cite{Zhou11}. Then, controlled phase-shift operations are applied to the modes $1$ and $2$ if the mode $3$ is again vertically polarized. Note that this process can be even deterministically implemented in superconducting qubits \cite{V13}.
Finally, by selecting out a click event on mode $3$, we can sort out the polarization-DOF entanglement in the same coherent state mode, 
$(|H\rangle_{A,1}|V\rangle_{A,2}-|V\rangle_{A,1}|H\rangle_{A,2})/\sqrt{2}$.

What if the complicated procedure of Fig. 3 is not perfectly set to operate? One of imperfections of experimental results is that optical components are not set to the correct values perfectly \cite{Zhou11}. Here we consider the imperfections of displacement operations or controlled operations. In Fig. 3 (a), given that the displacement operations do not produce  appropriate displacement amplitudes as $\hat{D}_{3,H}(B)$ and $\hat{D}_{4,V}(B)$ ($B$ is not equal to $A$), then the control mode 3 of Eq. (6) is represented in terms of horizontal and vertical polarizations simultaneously. Thus, without perfect displacement operations, one cannot control the target modes 1 and 2 with either of the polarizations in mode 3. In Fig. 3 (b), given that the vertical polarization of the control mode 3 is not properly identified, the polarizations of the modes 1 and 2 cannot be flipped perfectly. Also, the controlled phase-shift operations cannot control the phases of the modes 1 and 2 perfectly. Thus, the imperfection of the controlled operations produce a non-maximal polarization entangled state.

\section {Entanglement Observation}

We can observe entanglement in each DOF by means of different types of quantum information protocols. Parity entanglement can be verified by the CHSH-Bell type inequality which utilizes displaced parity measurement \cite{BW99}. 
It violates the inequality up to Tsirelson's bound $2\sqrt{2}$ with increasing average photon number \cite{WJK02}. 
Since the state 
\begin{equation}
|even\rangle_{1}|odd\rangle_{2}-|odd\rangle_{1}|even\rangle_{2}
\label{eq:entangledstate3}
\end{equation}
has faster oscillating amplitude in phase space with increasing average photon number,
one has more possibility of violating the inequality.

Let us conceive another application of the entangled state \eqref{eq:entangledstate3}. Applying local displacement operation $\hat{D}(\alpha)$ to each mode, we obtain a NOON-type coherent state without changing the degree of entanglement. It is known to provide the Heisenberg limit in local quantum phase estimation, specifically using photon number resolving detection in a Mach-Zehnder interferometer \cite{LLLN15}. We observe that the phase sensitivity increases with the increasing average photon number, and it goes down to the shot-noise limit with the decreasing of its entanglement. 

Next, polarization entanglement can be attested by interaction-free measurement (IFM) in the Mach-Zehnder (MZ) interferometer, which is also known as the Elitzur-Vaidman bomb tester \cite{EV93}. Injecting a single photon into the MZ interferometer without any obstacle (a bomb) in it, the quantum interference mechanism leads to detection only at the first detector. If a bomb is put in one of the internal arms of the MZ interferometer, the interference is disturbed and the photon is now either detected by the first or the second detector, or it hits the bomb---then it explodes. The efficiency of this detection strategy is formulated as $\eta=P_{IFM}/(P_{bomb}+P_{IFM})$, where $P_{IFM}$ is the probability of detecting the presence of the bomb and $P_{bomb}$ the probability of bomb explosion. For a single run, the efficiency of detection without explosion is as large as up to $50\%$.

\begin{figure}
\centerline{\scalebox{0.32}{\includegraphics[angle=0]{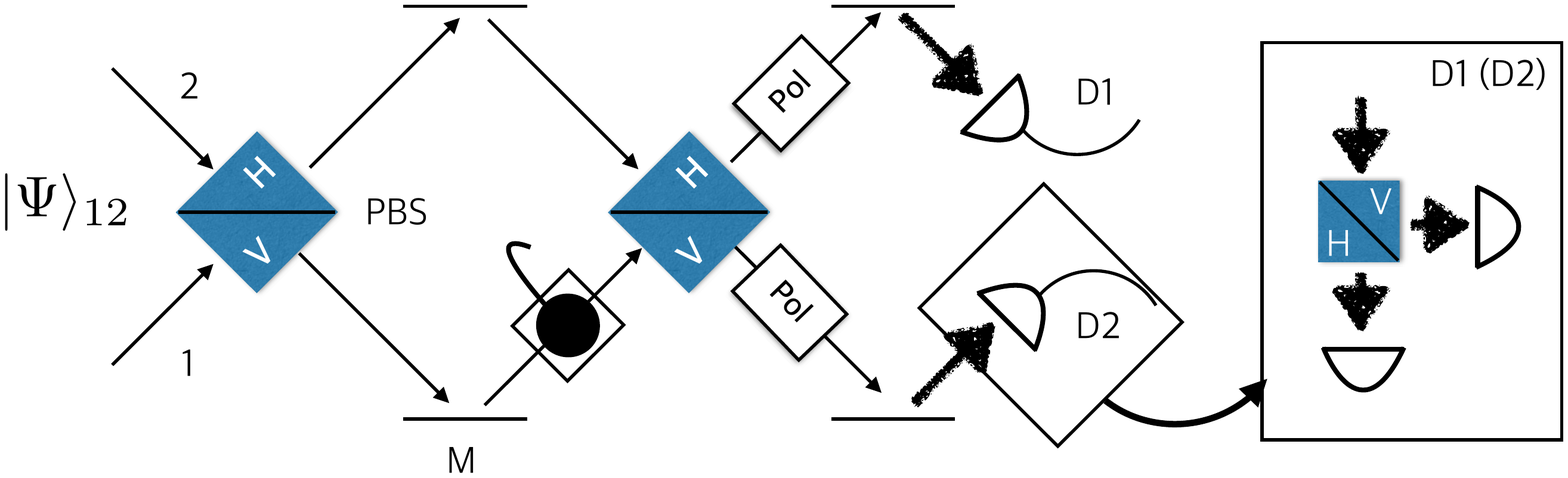}}}
\vspace{-1.6in}
\caption{Observing indistinguishability of polarization entangled photons by interaction-free measurement (IFM), where 
$|\Psi\rangle_{12}=\frac{1}{\sqrt{2}}(|H\rangle_1|V\rangle_2+|V\rangle_1|H\rangle_2)$. Pol is a 45-degree polarizer ($|H\rangle\rightarrow \frac{1}{\sqrt{2}}(|H\rangle+|V\rangle)$ and $|V\rangle\rightarrow \frac{1}{\sqrt{2}}(|V\rangle-|H\rangle)$).  PBS is a polarizing beam splitter.
D1 (D2) is composed of a PBS and two on-off detectors, and we can discriminate four different events [$(H,H),~(V,V),~(H,V),~(V,H)$] by the two detection setups.
}
\label{fig:fig4}
\end{figure}

In Fig. \ref{fig:fig4}, we consider an IFM setup that is composed of the MZ interferometer using PBSs. Note that a single-photon based IFM setup uses beam splitters instead. 
Assuming that our polarization entangled state has the constraint of $|A|^2=1$, we can take our entangled state $(|H\rangle_{A,1}|V\rangle_{A,2}-|V\rangle_{A,1}|H\rangle_{A,2})/\sqrt{2}$ as a two-particle entangled state formula $(|H\rangle_1|V\rangle_2-|V\rangle_1|H\rangle_2)/\sqrt{2}$. 
Injecting the polarization entangled state $|\Psi\rangle_{12}=(|H\rangle_{1}|V\rangle_{2}+|V\rangle_{1}|H\rangle_{2})/\sqrt{2}$ into the MZ interferometer without a bomb, the state is given by $|\Psi\rangle_{12}$ after the 2nd polarizing beam splitter. Then, applying a 45-degree polarizer to each mode, we obtain the output state $(|V\rangle_{1}|V\rangle_{2}-|H\rangle_{1}|H\rangle_{2})/\sqrt{2}$ and observe the same polarization states on each detector simultaneously. 
If, however, there is a bomb in one of the arms, the output state is given by $(|V\rangle_{1}|V\rangle_{2}-|H\rangle_{1}|H\rangle_{2}+|H\rangle_1|V\rangle_2-|V\rangle_1|H\rangle_2)/2\sqrt{2}$. 
As shown in Fig. 4, we can discriminate the four different events by a combination of PBSs and on-off detectors.
Then, we are to observe the different polarization states on both detectors simultaneously with $25\%$ probability. Thus, the polarization entangled state attains $33.3\%$ efficiency of the IFM since the probability of detecting the presence of the bomb is a half to that of bomb explosion. If it is not maximally entangled, there are also four different detection events in no bomb scenario. We cannot discriminate the two scenarios of no bomb and bomb, such that the efficiency of detection without explosion is equal to zero.

\section{Alternative implementation}
We can also consider an alternative implementation which takes into account a squeezed vacuum (SV) state ($|S\rangle\equiv \hat{S}(r)|0\rangle$) instead of a coherent state. In state generation stage, SV state case is more feasible than the coherent state one. However, the SV state case is more difficult to access entanglement than the coherent state one. 

 First, an entangled state based on squeezed vacuum (SV) state is produced by applying a coherent superposition operation of photon subtractions to two single-mode SV states. Its output state is derived as
\begin{eqnarray}
(\hat{a}_1+\hat{a}_2)|S\rangle_1|S\rangle_2=\hat{a}_1|S\rangle_1|S\rangle_2+|S\rangle_1\hat{a}_2|S\rangle_2,
\end{eqnarray}
where the photon-subtracted SV state $\hat{a}|S\rangle$  is a superposition of odd number states and the SV state $|S\rangle$  is a superposition of even number states. Then, sequentially applying polarizers ( mode 1 $\rightarrow$ H, mode 2 $\rightarrow$ V) and a PBS to the output state, we obtain an entangled state which has capability of \emph{entanglement duality},
\begin{eqnarray}
\frac{1}{\sqrt{2}\sinh{r}}(\hat{a}_H|S\rangle_H|S\rangle_V+|S\rangle_H\hat{a}_V|S\rangle_V),
\label{eq:svpolarization}
\end{eqnarray}
where $r$ is a squeezing parameter.
Note that the entangled state of Eq. (9) has been implemented in optical frequency range in order to mimic entangled coherent states \cite{OFTG09}.

 It is expected that it is easier to generate the entangled state of Eq. (4) which requires a preliminary stage of preparing an odd coherent state. Furthermore, replacing $\hat{a}_1+\hat{a}_2$  with a modified coherent superposition operation $\sqrt{T}\hat{a}_1+\sqrt{1-T}\hat{a}_2$ \cite{LN10}, we can control the amount of indistinguishability of the entangled polarization-SV state. 
It is implemented by adjusting the transmittance of a beam splitter \cite{LN10}.
However the indistinguishability in Eq. (4) is not implementable simply by controlling a beam splitting parameter. 
The beam splitting parameter just changes the amplitude of the coherent state basis.

In spite of the simple state preparation, it is not simple to access the corresponding DOF to observe polarization entanglement. The SV state case requires an additional process of distinguishing even and odd number states without destroying the states.
The details of the process are given in Appendix A. It is due to the property that a single-mode squeezing operator produces a two-mode squeezing operation when we impinge a single-mode squeezed vacuum state on a beam splitter. It makes the access of entanglement more complicated, compared to the coherent state basis.

Provided we can access the entanglement, we can test its duality by the same quantum information protocols as in the coherent state case. Moreover, by applying a single-mode  anti-squeezing operation $\hat{S}(-r)$ to each mode, we obtain the single-photon entangled state $\frac{1}{\sqrt{2}}(|1\rangle_1|0\rangle_2+|0\rangle_1|1\rangle_2)$, without changing the degree of  entanglement. This can be used for  the quantum teleportation protocol based on single-photon entanglement \cite{LK00}.

\section{Summary and Discussion}

We have proposed a scheme to test the indistinguishability of macroscopic entangled states of light. It has been shown that the duality in entanglement  between polarization and parity DOFs  can be accessed under current technology.  Then we have mentioned that  parity entanglement and polarization entanglement can be verified by CHSH-Bell type inequality and interaction-free measurement, respectively. Furthermore, we  have proposed an alternative implementation scheme using a squeezed vacuum state.

Here, we have considered many particles in a bipartite system. It would be interesting to extend our scenario to many particles in a multi-partite system.  In order to observe full indistinguishability of the multi-partite system, we need more degrees of freedom for each party.  
Full indistinguishability is confirmed by full interchangeability of degrees of freedom, which is achieved when the number of degrees of freedom is equal to the number of the parties.
A candidate is time-frequency modes in optical frequency combs \cite{R16} and multi-headed coherent states \cite{R10,L13,K15,LLNK15}. 
From the squeezed vacuum state case, we expect to find a way of discriminating even and odd numbers without destroying states. Furthermore, we could get an idea of a generalized parity measurement, i.e., modulo operation.

\begin{acknowledgments}
SYL thanks Dr Changsuk Noh and Prof Hyunseok Jeong for useful comments.
SYL, CWL, and JK were partly supported by the IT R$\&$D program of MOTIE/KEIT [$10043464$]. PK was supported by the National Science Centre in Poland through the NCN Grant No. 2014/14/E/ST2/00585.
\end{acknowledgments}

\appendix
\section{Observing polarization entanglement in  squeezed vacuum (SV) state case}
 We need to attach (on mode $3$) an H-polarized single-photon state to the state in Eq. \eqref{eq:svpolarization},
\begin{eqnarray}
(\hat{a}_{H,1}|S\rangle_{H,1}|S\rangle_{V,2}+|S\rangle_{H,1}\hat{a}_{V,2}|S\rangle_{V,2})|H\rangle_3,\nonumber
\end{eqnarray}
where the state is unnormalized.
If the mode $2$ consists of odd number states, we  conditionally flip the polarization of the mode $3$ as follows,
\begin{eqnarray}
\hat{a}_{H,1}|S\rangle_{H,1}|S\rangle_{V,2}|H\rangle_3+|S\rangle_{H,1}\hat{a}_{V,2}|S\rangle_{V,2}|V\rangle_3.\nonumber
\end{eqnarray}
Next, if the mode $3$ is vertically polarized, we apply two controlled NOT gates [as in Fig. \ref{fig:fig3}(b)] to conditionally flip the polarization of the modes 1 and 2,
\begin{eqnarray}
\hat{a}_{H,1}|S\rangle_{H,1}|S\rangle_{V,2}|H\rangle_3+|S\rangle_{V,1}\hat{a}_{H,2}|S\rangle_{H,2}|V\rangle_3.\nonumber
\end{eqnarray}
Applying a controlled-SWAP operation under the control mode $3$, the output state is   obtained
\begin{eqnarray}
\hat{a}_{H,1}|S\rangle_{H,1}|S\rangle_{V,2}|H\rangle_3+\hat{a}_{V,1}|S\rangle_{V,1}|S\rangle_{H,2}|V\rangle_3,\nonumber
\end{eqnarray}
where the controlled-SWAP operation can be implemented by an optical Fredkin gate \cite{GC01,P16}. Finally, applying photon subtraction operation to mode $2$ and detecting $45$-degree polarization ($|H\rangle+|V\rangle$) on mode $3$, we obtain the polarization entangled state on the same photon-subtracted SV state mode, $|H\rangle_1|V\rangle_2+|V\rangle_1|H\rangle_2$.

\end{document}